# The Effect of Nanomagnet Geometry on Reliability, Energy Dissipation and Clock Speed in Strain-Clocked Dipole-Coupled Nanomagnetic Logic

Md Mamun Al-Rashid, Dhritiman Bhattacharya, Supriyo Bandyopadhyay, Fellow, IEEE
Jayasimha Atulasimha, Senior Member, IEEE.

*Abstract*—Strain-clocked dipole-coupled nanomagnetic logic is an energy-efficient Boolean logic paradigm whose progress has been stymied by its propensity for high error rates. In an effort to mitigate this problem, we have studied the effect of nanomagnet geometry on error rates, focusing on elliptical and cylindrical geometries. We had previously reported that in elliptical nanomagnets, the out-of-plane excursion of the magnetization vector during switching creates a precessional torque that plays a dual role – it speeds up the switching but is also responsible for the high switching error probability. The absence of this torque in cylindrical magnets portends lower error rates, but our simulations show that the error rate actually does not improve significantly compared to elliptical magnets while the switching becomes unacceptably slow. Here, we show that dipole coupled nanomagnetic logic employing elliptical nanomagnets can offer relatively high reliability for nanomagnetic logic (switching error probability < $10^{-8}$), moderate clock speed (~ 100 MHz) and 2-3 orders of magnitude energy saving compared to CMOS devices, provided the shape anisotropy energy barrier of the nanomagnet is increased to at least ~5.5 eV to allow engineering a stronger dipole coupling between neighboring nanomagnets.

## I. Introduction

Dipole-coupled nanomagnetic logic (DC-NML) implemented with single-domain elliptical nanomagnet "switches" whose two stable magnetization orientations encode binary bit information [1, 2] is a popular paradigm for energy-efficient Boolean computing. It is also attractive from an architectural perspective; the non-volatility of the magnets allows the same device to act as both processor and memory, thereby obviating the need for processor/memory partition.

Unfortunately, not all renditions of DC-NML are necessarily energy-efficient. If the single domain [3] magnets are switched between the stable orientations with a magnetic field [4], or a spin polarized current acting as a clocking agent [5], the associated energy dissipation in the clock becomes so large that it offsets any energy-advantage of DC-NML. Recent proposals have therefore explored ways of drastically reducing the clock dissipation by using nanomagnets with perpendicular anisotropy [6], multiferroic nanomagnets whose magnetizations are switched with strain [7, 8] and Spin Hall Effect (SHE) to inject spin polarized current into a nanomagnet for switching [9].

The Achilles' heel of strain-clocked DC-NML is its poor reliability due to high switching error rates at room temperature [10-14]. In this paper, we explore ways of mitigating the poor reliability, particularly through the use of appropriate geometry of the nanomagnets, and identify the metrics that have to be sacrificed to attain increased robustness. For this purpose, we compare two renditions of strain-clocked DC-NML that are differentiated by the geometrical shapes of the nanomagnets used as the binary switches: (1) the nanomagnets are *cylindrical pillars* with two stable magnetization orientations along the two (mutually antiparallel) orientations collinear with the cylinder's axis, and (2) the nanomagnets are *elliptical discs* [7] (major and minor axes of the ellipse much larger than the thickness) and the two stable magnetization directions are along the major axis of the ellipse.

DC-NML of the latter variety (elliptical discs) is error-prone owing to the effect of the magnet geometry on switching dynamics. This can be understood by looking at the illustration in Fig. 1 where the magnetization vector is represented in spherical coordinates with polar angle $\theta$ and azimuthal angle $\phi$. The polar angle $\theta$ is a measure of the out-of-plane excursion of the magnetization vector; $\theta < 90^0$ and $\theta > 90^0$ respectively imply that the magnetization is above/below the plane of the magnet. Whenever the magnetization vector leaves the magnet's plane during switching, its out-of-plane component produces a demagnetization field in the out-of-plane direction which generates a torque on the magnetization which either assists or hinders switching depending on whether the magnetization vector is above or below the plane of the magnet (x-y plane) [15]. Failure to switch will constitute an "error". If there are neighboring magnets that interact with the test magnet via dipole coupling, then the resulting dipole field can be utilized to counter the hindering torque at least partially and improve the switching error rate. However, the efficacy of this strategy may be limited by geometric constraints such the minimum allowable separation between neighboring magnets (which determines the dipole coupling strength) and the spread in the out-of-plane excursion of the magnetization vector at the operating temperature. Limiting the out-of-plane excursion by judicious choice of nanomagnet geometry therefore appears to be an appropriate route to reducing the frequency of error or probability of error.

$^S$ubmitted on April 17, 2015. This work is supported by the National Science Foundation (NSF) under CAREER grant CCF-1253370.

M. Al-Rashid and J. Atulasimha are with both the Department of Mechanical and Nuclear Engineering and the Department of Electrical and Computer Engineering, Virginia Commonwealth University, Richmond, VA 23284 USA (e-mail: alrashidmm@vcu.edu, jatulasimha@-vcu.edu).

D. Bhattacharya is with the Department of Mechanical and Nuclear Engineering Virginia Commonwealth University, Richmond, VA23284 USA (e-mail: bhattacharyad@vcu.edu).

S. Bandyopadhyay is with the Department of Electrical and Computer Engineering, Virginia Commonwealth University, Richmond, VA23284 USA (e-mail: sbandy@vcu.edu)

Cylindrical nanomagnets shown in Fig. 1(a) and Fig. 1(b) have a geometry that can quench or eliminate the offending precessional torque. We can ensure that the cylinder's axis is the easy magnetization direction by making the ratio of the cylinder's height to diameter larger than 0.91 [16]. When the magnetization is switched from the "up" ($\theta=0°$) to the "down"($\theta = 180°$) state, there is no "out-of-plane" or "in-plane" direction perpendicular to the cylinder axis since the cross-section is circular and therefore perfectly symmetric in the plane perpendicular to the cylinder's axis. While this could potentially reduce switching error by eliminating the torque associated with the out-of-plane excursion, the downside is that the absence of this torque would make switching slow because the magnetization has to switch via the damped mode torque alone since the (much stronger) precessional mode torque associated with out-of-plane excursion no longer exists. This makes the comparison between the switching dynamics of the two types of NML, and the associated switching errors and switching delay, an interesting problem.

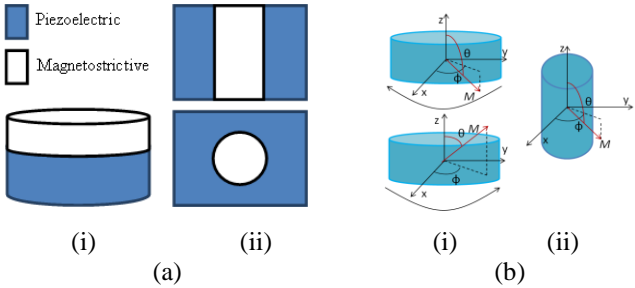

(a)          (b)

Fig. 1. a) (i) An elliptical disc multiferroic nanomagnet comprising a piezoelectric layer and a magnetostrictive layer that are elastically coupled. (ii) A cylindrical nanomagnet embedded in a piezoelectric matrix – cross-sectional side view (top) and top view (bottom). b) Magnetization orientation *M* in: (i) an elliptical and (ii) a cylindrical nanomagnet. For the ellipse, we show the magnetization vector below and above the magnet's plane and the corresponding direction of the precessional torque (clockwise and counterclockwise) resulting from the out-of-plane excursion of the magnetization vector.

Strain-clocked *elliptical* nanomagnets are implemented by delineating a single domain magnetostrictive layer on top of a piezoelectric layer in the manner of [7] to form a two-phase multiferroic nanomagnet, while strain-clocked *cylindrical* nanomagnets are implemented by embedding the ferromagnetic cylinders within a piezoelectric material in the manner of [17, 18]. The nanomagnets can be switched by applying a small voltage/electric field to the piezoelectric layer that produces a strain which is transferred to the elastically coupled magnetostrictive ferromagnet and rotates its magnetization [7, 17]. In this paper, simulations are performed for elliptical disks that are 58 nm in length (major axis), 40 nm in width (minor axis) and 12 nm in thickness, while the cylindrical nanomagnets are 35 nm tall and have a cross-sectional diameter of 28 nm. Therefore, they have similar volumes (21865 and 21551 nm$^3$ respectively) that are within 3% of each other. We understand that tolerances of few nanometers in lateral dimensions may be hard to obtain, but this design is primarily intended for a theoretical comparison between two geometries and it is vitally important to ensure that the volume and shape anisotropy barriers are as close as possible to make a fair comparison. These magnets have been designed such that the shape anisotropy energy barrier is approximately ~5.5 eV or ~220 kT at room temperature (k is the Boltzmann constant and T is the absolute temperature). Ferromagnets of these dimensions are typically single domain [3]. In equilibrium, the magnetization vectors of these magnets are directed along the major axis (easy axis) of the ellipse (y-axis) and the axis of the cylinder (z-axis), respectively. Thermal noise will cause the magnetization to fluctuate around these positions, but these positions are the most probable orientations.

## II. THEORY

The switching dynamics in both geometries is simulated by solving the Landau–Lifshitz–Gilbert (LLG) equation [19, 21] under the macrospin assumption that is reasonable for nanomagnets of the chosen dimensions [3]. We consider a pair of nanomagnets separated along the x-axis (their hard axis) with center-to-center distance of *R*. We then study the switching dynamics of the second nanomagnet (one on the right) under the dipole coupling influence of the first (one on the left) while "clocking" the second nanomagnet with (locally applied) uniaxial compressive stress.

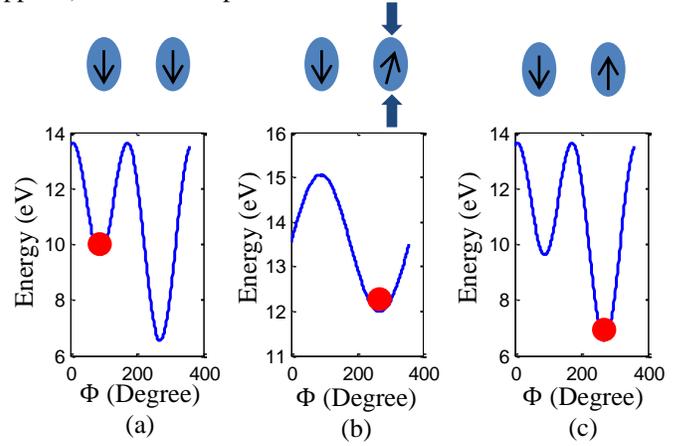

(a)          (b)          (c)

Fig. 2. Strain-clocking of a dipole coupled pair of ellipsoidal nanomagnets to implement a Boolean NOT gate. The left nanomagnet's magnetization encodes the input bit and the right nanomagnet's magnetization encodes the output bit. The right nanomagnet is stressed locally to make its magnetization rotate by ~90$^0$ and align with the minor axis. Upon withdrawal of the stress, the right nanomagnet's magnetization preferentially assumes an orientation anti-parallel to that of the left's because of dipole coupling with the left neighbor, thereby realizing the Boolean NOT operation (output is the logic complement of the input): clocking sequence (top) and energy profile of the right magnet (bottom). The energy profile shows the total potential energy vs. in-plane magnetization orientation of the right magnet: (a) before application of stress, (b) after application of critical stress and (c) after stress withdrawal. The critical stress is the stress that makes the stress anisotropy potential energy barrier become equal to the shape anisotropy energy barrier of the right nanomagnet. The red ball in the energy profiles depicts the in-plane magnetization orientation of the magnetization vector of the right magnet in different conditions. Similar chain of events occurs during the switching of a cylindrical nanomagnet where the angle of rotation is the polar angle $\theta$.

Note that the magnetization orientation for the second nanomagnet also affects the magnetization orientation of the first through the dipole coupling. However, since the shape anisotropy of the first nanomagnet is rather large and no stress is applied on it to lower its shape anisotropy energy barrier of ~220 kT at room temperature (~5.5 eV), the perturbation in its magnetization orientation is small when the second magnet's magnetization rotates. Thus, though the magnets are of identical dimensions, unidirectional information flow from the first to the second magnet is enforced by the stress/strain that clocks the second magnet selectively. This clocking can be extended to a chain of nanomagnets in the manner of reference 7 to carry out Bennett clocking and unidirectional flow of logic information.

The magnetization dynamics of a single-domain magnetostrictive nanomagnet is governed by the Landau-Lifshitz-Gilbert (LLG) equation:

$$\frac{d\vec{M}(t)}{dt} = -\gamma \vec{M}(t) \times \vec{H}_{eff}(t) - \frac{\alpha\gamma}{M_S}\left[\vec{M}(t) \times \left(\vec{M}(t) \times \vec{H}_{eff}(t)\right)\right], \quad (1)$$

where $\vec{H}_{eff}(t)$ is the effective magnetic field felt by the nanomagnet due to stress, shape-anisotropy and dipole coupling with neighbor(s). It is given by the derivative of the total potential energy $E(t)$ with respect to the magnetization $\vec{M}(t)$:

$$\vec{H}_{eff}(t) = -\frac{1}{\mu_0 \Omega} \frac{\delta E(t)}{\delta \vec{M}(t)}, \quad (2)$$

where $M_S$ is the saturation magnetization of the nanomagnet, $\mu_0$ is the permeability of vaccum, $\gamma$ is the gyromagnetic ratio, $\Omega$ is the volume of the nanomagnet, and $\alpha$ is the Gilbert damping constant. The first term on the right hand side of equation (1) relates to the precessional torque and the second term to the damped-mode torque.

The total energy $E(t)$ in (2) is given by:

$$E(t) = E_{dipole}(t) + E_{stress\,anisotropy}(t) + E_{shape\,anisotropy}(t), \quad (3)$$

where $E_{dipole}(t)$ is the dipole coupling energy due to interaction between the two nanomagnets, $E_{shape\,anisotropy}(t)$ is the shape anisotropy energy due to the elliptical or cylindrical shape of the nanomagnet, and $E_{stress\,anisotropy}(t)$ is the stress anisotropy energy due to the stress generated in the nanomagnet. The analytical expression for each of these energies, for both the elliptical and cylindrical nanomagnet geometries, can be found in Appendix A.

The effect of thermal noise is incorporated by adding an equivalent field $\vec{H}_{thermal}(t)$ to the total effective field [11, 12, 15, 20-22]:

$$\vec{H}_{eff}(t) = -\frac{1}{\mu_0 \Omega} \frac{\delta E(t)}{\delta \vec{M}(t)} + \vec{H}_{thermal}(t). \quad (4)$$

It is modeled as a random field as described in [21].

$$\vec{H}_{thermal}(t) = \sqrt{\frac{2kT\alpha}{\mu_0 M_S \gamma \Omega \Delta t}}\left(\vec{G}(t)\right), \quad (5)$$

where $\vec{G}(t)$ is an independent Gaussian distribution with zero mean and unit variance in each Cartesian coordinate axis.

Equation (1) can be simplified by normalizing the magnetization vector with respect to the saturation magnetization $M_s$.

$$\vec{m} = \vec{M}/M_s, \quad m_x^2 + m_y^2 + m_z^2 = 1 \quad (6)$$

where $m_x$, $m_y$ and $m_z$ are the x, y and z component of the normalized magnetization vector $\vec{m}$ respectively that are given by:

$$m_x(t) = \sin\theta(t)\cos\phi(t), \quad m_y(t) = \sin\theta(t)\sin\phi(t), \quad m_z(t) = \cos\theta(t) \quad (7)$$

Using these relations, the vector LLG equation can be decomposed into two coupled scalar equations that describe the time evolution of the azimuthal ($\phi$) and polar ($\theta$) angles of the magnetization vector:

$$\begin{aligned}\frac{d\theta(t)}{dt} = &\frac{1}{\cos\theta(t)}[\cos\phi(t)\{-\gamma(H_{eff-z}(t)m_y(t) - H_{eff-y}(t)m_z(t))\\&-\alpha\gamma(H_{eff-y}(t)m_x(t)m_y(t) - H_{eff-x}m_y^2(t) - H_{eff-x}(t)m_z^2(t)\\&+H_{eff-z}(t)m_x(t)m_z(t))\} + \sin\phi(t)\{-\gamma(H_{eff-x}(t)m_z(t)\\&-H_{eff-z}(t)m_x(t)) - \alpha\gamma(H_{eff-z}(t)m_y(t)m_z(t)\\&-H_{eff-y}(t)m_z^2(t) - H_{eff-y}(t)m_x^2(t) + H_{eff-x}(t)m_x(t)m_y(t))\}]\end{aligned} \quad (8)$$

$$\begin{aligned}\frac{d\phi(t)}{dt} = &\frac{1}{\sin\theta(t)}[\sin\phi(t)\{-\gamma(H_{eff-z}(t)m_y(t) - H_{eff-y}(t)m_z(t))\\&-\alpha\gamma(H_{eff-y}(t)m_x(t)m_y(t) - H_{eff-x}m_y^2(t) - H_{eff-x}(t)m_z^2(t)\\&+H_{eff-z}(t)m_x(t)m_z(t))\} + \cos\phi(t)\{-\gamma(H_{eff-x}(t)m_z(t)\\&-H_{eff-z}(t)m_x(t)) - \alpha\gamma(H_{eff-z}(t)m_y(t)m_z(t)\\&-H_{eff-y}(t)m_z^2(t) - H_{eff-y}(t)m_x^2(t) + H_{eff-x}(t)m_x(t)m_y(t))\}]\end{aligned} \quad (9)$$

Here $H_{eff-x}$, $H_{eff-y}$ and $H_{eff-z}$ are the x, y and z components of the effective magnetic field $\vec{H}_{eff}$ that are evaluated using (4) and (5). Analytical expressions for the effective field for both the elliptical and cylindrical nanomagnet geometries can be found in Appendix A. In both cases, stress is only applied to the second nanomagnet in the chain.

In all simulations, the magnetostrictive ferromagnetic material is Terfenol-D which has the following parameters-

TABLE I
TERFENOL-D MATERIAL PROPERTIES

| Parameters | Values |
|---|---|
| Young's modulus (Y) | $8\times10^{10}$ Pa |
| Magnetostrictive coefficient $((3/2)\lambda_S)$ | $90\times10^{-5}$ |
| Saturation magnetization ($M_s$) | $8\times10^5$ A/m |
| Gilbert's damping constant ($\alpha$) | 0.1 |

### III. RESULTS AND DISCUSSION

We study the switching time as well as the switching error probability for both geometries for varying dipole strengths. An increase in dipole coupling energy (smaller separation between the nanomagnets) would produce a higher effective field and make the switching faster in both geometries. This corresponds to the steeper slope in the energy profile shown in Fig. 2(b). Further, stronger dipole coupling introduces a larger asymmetry in the potential profile shown in Fig. 2 that improves the probability of switching to the correct state, even in the presence of thermal noise. While the above behavior is expected for both geometries, the interesting question is how the two geometries differ with respect to switching speed and error. This is discussed next after briefly explaining the simulation conditions and procedures.

## A. Simulation conditions: stress application

A compressive stress exactly equal to the critical stress is applied in the elliptical (44.28 MPa) and cylindrical (45.85 MPa) nanomagnets (see the caption of Fig. 2 for definition of the term "critical stress"). Our previous work had shown that for a given dipole coupling, the switching probability is highest (error probability least) when the stress applied is the critical stress [10].

## B. Switching time estimate

The switching trajectories and the corresponding switching times are random in the presence of thermal noise. Because we are interested in the *difference* between the two geometries, we adopt the following strategy. We perform stochastic LLG simulations in the presence of thermal noise to determine the thermal distribution of the magnetization vector around a stable orientation, and randomly pick a starting point from this distribution. The stress pulse is applied to kick the magnetization out of its initial stable orientation around $\phi = 90^0$ and set it off towards the intended final stable orientation around $\phi = 270^0$. We simulate the temporal evolution of the magnetization orientation from (8) and (9), and determine the time taken for the magnetization orientation to reach close to $\phi=270°$ (the switching is deemed to have occurred if the deviation of the final value of $\phi$ from $\phi=270°$ is within 1°). This process is repeated to generate different switching trajectories. The fraction of the trajectories that fail to reach close to $\phi = 270^0$ is the error probability. A similar methodology is used for the cylinder case. In both cases, switching occurs with highest probability because we use critical stress that just erodes the shape anisotropy barrier and does not force the magnetization to orient close to the hard axis, and thereafter makes the magnetization switch because of dipole coupling with the left neighbor (which prefers anti-ferromagnetic ordering). The mean switching time is calculated by averaging over the successful trajectories. We also find the longest switching time (from the slowest trajectory) to assess the worst case scenario. The energy dissipation is calculated in the manner of [22]. It includes the internal dissipation in the magnet due to Gilbert damping and the $(1/2)CV^2$ dissipation associated with charging the capacitor $C$ formed by the piezoelectric layer, with $V$ being the voltage needed to produce the electric fields in the piezoelectric to generate the stress.

In case of elliptic nanomagnets, the capacitance $C$ is estimated assuming that two square electrodes of side ~50 nm are used to apply the voltage over a PZT layer of thickness ~50 nm in the manner of [23]. For applying stress to the cylindrical nanomagnet in the manner of [17], we assume, the PZT matrix is ~ 70 nm thick and the capacitor plate is square with side dimension of ~70 nm.

## C. Comparison between the elliptical and cylindrical geometries in terms of switching time or switching speed

Fig. 3(a) and Fig. 3(b) show the switching times for elliptical and cylindrical geometries assuming comparable dipole coupling strengths. As expected, increased dipole coupling decreases the switching time in both cases. However, at any given dipole coupling strength, the switching time is ~10 to 50 times (1-2 orders of magnitude) *longer* for the cylindrical geometry compared to the elliptical one. *This highlights the critical role played by the switching geometry* in determining the switching speed and hence, ultimately, the clock speed in DC-NML.

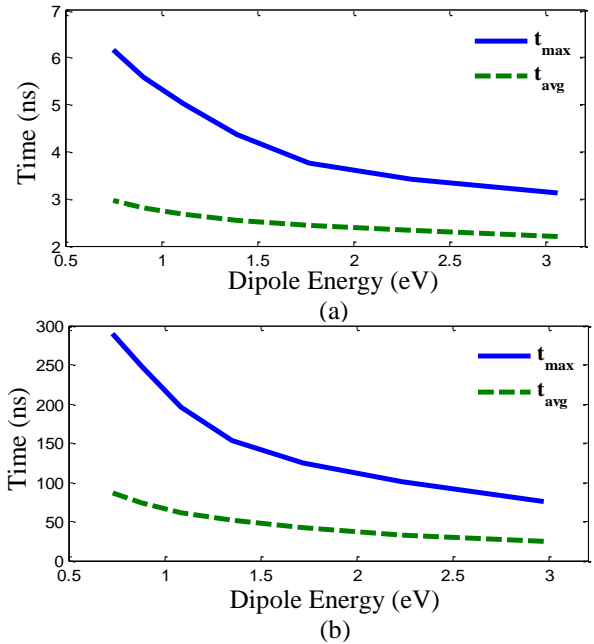

Fig. 3. Switching time vs. dipole energy (determined by the center-to-center separation between neighboring nanomagnets) for (a) elliptical nanomagnets and (b) cylindrical nanomagnets.

In the case of the elliptical magnet, when the magnetization rotates anticlockwise (or clockwise), the magnetization lifts out of the plane (or dips below the plane) [15, 20]. This produces a magnetization component in the positive and negative z-direction, respectively, that leads to an effective field in the negative or positive direction. This z-directed field increases the anticlockwise (or clockwise) torque, thereby increasing the speed of the switching.

However, in the case of the cylindrical geometry, as the magnetization switches from $\theta \sim 0^0$ to $\theta \sim 180^0$, the $\phi$-component of motion does not produce any additional torque since the geometry is completely symmetric with respect to $\phi$. Hence, the switching is primarily via the damped mode torque (unlike the elliptical geometry where the precessional mode torque plays a significant role in the switching process). Since most materials have a small Gilbert damping factor $\alpha$, the damped mode torque is usually far *weaker* than the precessional mode torque. This explains the extremely slow switching times in the cylindrical geometry and the difference with the elliptical geometry.

## D. Switching error estimate

The switching error probabilities in Fig. 4(a) and Fig. 4(b) were estimated by performing stochastic LLG simulations as described. The simulation was first run for 1 ns without applying any stress and the distribution of the magnetization orientation around the stable easy direction was obtained. Next, a switching trajectory was generated by solving (8) and (9). The starting point of this trajectory (at time t = 0) was picked from the distribution generated in the previous step with the corresponding weight. Thereafter, the stress was ramped up to the critical stress value for 1 ns, held for a period of time as described in the legends of Fig. 4 and then removed in a 1 ns downward ramp. The system was given ~ 1 ns (ellipse) and ~27 ns (for the cylinder) to come to a steady state. The relaxation time was determined by the time it took all the magnetization trajectories to end up in one of the stable states. The fraction of the number of trajectories that had not switched to the correct state constituted the switching error probability. For most cases, 100,000 trajectories were simulated at 300 K. However, in cases where we report error probabilities of ~$10^{-6}$, $10^{-7}$ and $10^{-8}$, the number of trajectories simulated was 1 million, 10 million and 100 million, respectively. Because simulation of so many trajectories is time consuming, such simulations were limited to a few cases where the dipole coupling strength was extremely high. The 1-100 million simulations cases were performed only on the elliptical geometry as it is computationally more tractable to do these simulations over a switching time ~10 ns as opposed to ~several 100 ns needed for the cylindrical geometry.

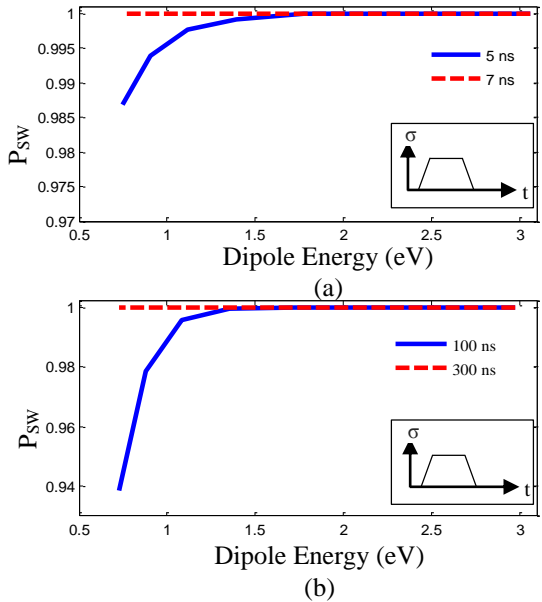

Fig. 4. Switching probability ($P_{SW}$) vs. dipole coupling energy (or equivalently center to center separation between neighboring nanomagnets) for (a) elliptical nanomagnets and (b) cylindrical nanomagnets. The results are shown for two different switching times. Initial time before application of stress, stress ramp up and stress ramp down times are fixed at 1 ns each. Final relaxation time, after stress in ramped down, is 1 ns for ellipse and 27 ns for the cylinder. The stress hold times are 1 and 3 ns for the two ellipse cases and 70 ns and 270 ns for the two cylinder cases. The total time (ramp up, hold and ramp down, relaxation time) is indicated on the figure legends.

*E. Comparison between the elliptical and cylindrical geometries in terms of switching error*

Fig. 4(a) and Fig. 4(b) respectively show the dynamic switching error vs. dipole coupling energy (which is ultimately the internal energy dissipated) for the elliptical and circular geometry. Despite the absence of the torque due to "out-of-plane" magnetization distribution in the cylindrical geometry, the switching error is *not* any better than the elliptical case where the detrimental effects of the "out-of-plane" magnetization distribution is successfully countered by strong dipole coupling. The torque produced by the out-of-plane excursion of the magnetization orientation significantly shortens the switching time in the elliptical geometry but does not increase the switching error probability in the range of dipole energies and error rates we study. If the dipole coupling strength had been weaker, the elliptical geometry would surely have been more error-prone than the cylindrical geometry because of the effect of the out-of-plane magnetization distribution, but in the limit of strong dipole coupling, the effect of the out-of-plane magnetization distribution is diminished to the point where the difference between the two geometries become nearly imperceptible.

Clearly, stronger dipole coupling will reduce the error rates in dipole coupled nanomagnetic logic. However, the dipole coupling energy cannot be increased arbitrarily; it *must never* exceed the shape anisotropy energy barrier in the nanomagnets since that would then align their magnetizations along the minor axes of the ellipses (the line joining their centers) resulting in ferromagnetic ordering that does not implement the NOT logic functionality. Therefore, increasing the dipole coupling necessitates increasing the shape anisotropy energy barrier as well. For safe operation, the latter should be maintained at somewhat above the maximum dipole coupling energy. In our case, it was approximately ~220 kT (~ 5.5 eV).

IV. CONCLUSION

We found that increased dipole coupling strength results in lower error probability and faster switching, but obviously at the expense of higher energy dissipation since stronger dipole coupling causes larger dissipation [15]. We also found that we obtain comparable error probabilities with comparable energy dissipation but much faster switching speed for the elliptical geometry compared to the cylindrical geometry for the dipole coupling strengths we have considered. Thus, the elliptical geometry produces a very favorable energy-delay product for a given error rate, compared to the cylindrical geometry, as shown in Fig. 5(a). From the case with 100 million simulations, we determined that we can get an error probability < $10^{-8}$ in an elliptical magnet with an energy-delay product ~ $4.43 \times 10^{-26}$ J-s. Current CMOS devices have energy-delay product ~ $1.35 \times 10^{-25}$ J-s [24] and switching error probability < $10^{-12}$.

It is also very important to look at the error vs. energy dissipation plot (Fig. 5(b)). Here, the elliptical nanomagnet can switch with ~$10^{-8}$ or lower dynamic error probabilities at room temperature with very little energy dissipation (~8.87 aJ). To reduce the energy dissipation, the energy barrier could be lowered while simultaneously increasing the nanomagnet volume by making the aspect ratio (major axis/minor axis) of the ellipse smaller as long as the single domain approximation is still valid. This significantly reduces the stress required and therefore, the voltage that must be applied to clock the nanomagnet. "Elliptical nanomagnet-1" in Fig. 5(b) is one such design that would dissipate even less energy (~0.6 aJ) while dynamic switching error probability remains smaller than $10^{-8}$. Thus, these strain clocked NML switches dissipate 2 to 3 orders of magnitude less energy than a state-of-the-art CMOS switch which dissipates ~440 aJ. However, the CMOS switch is also less error prone with dynamic switching error probability typically $< 10^{-12}$.

In general, a CMOS switch may outperform dipole coupled nanomagnetic logic in switching speed and error rates, but it is usually much more dissipative and most importantly, it is volatile. In some niche applications such as embedded processors, where energy is a premium, $10^{-8}$ error probability can be tolerated and clock speeds ~ 100 MHz are sufficient. There, dipole coupled nanomagnetic computing schemes, clocked in an energy efficient manner (for example with strain), may steal a march over traditional CMOS-based implementations.

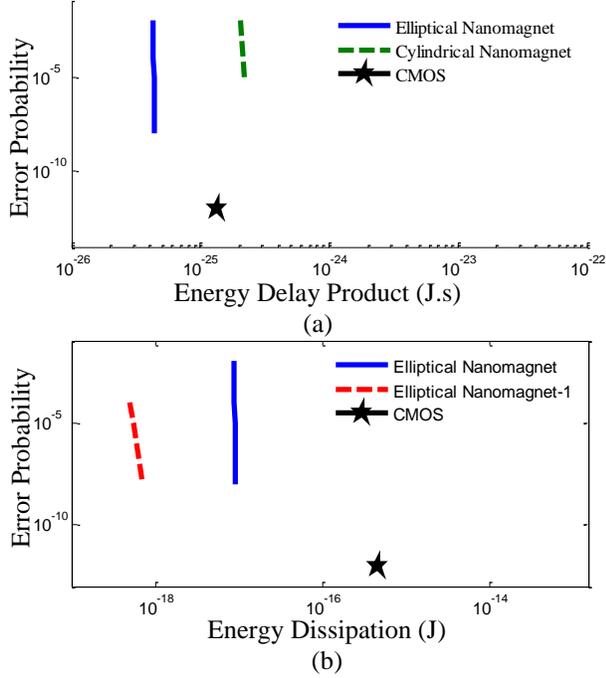

Fig. 5. (a) Comparison between elliptical and cylindrical geometries and CMOS: Error probability vs. energy-delay product. (b) Comparison between two different elliptical geometries and CMOS: Error probability vs. energy dissipation. Note: (i) Elliptical nanomagnet: major axis= 58 nm major axis, minor axis =40 nm and thickness = 12 nm (single domain approximation is good: see appendix: B). (ii) Elliptical nanomagnet-1: major axis= 110 nm major axis, minor axis =100 nm and thickness = 11nm (single domain behavior is still a good approximation). (iii). The CMOS energy-dissipation and the energy-delay product are taken from [24].

It should be noted that our theoretical error estimates assume that there are no fabrication defects such as variations in nanomagnet dimensions and misalignments between nanomagnets which can cause additional errors. These are not easily quantifiable and hence not addressed here. We focus only on intrinsic errors caused by thermal noise to estimate a theoretical limit on the reliability of dynamic switching.

## APPENDIX A

### I. EXPLANATION OF ENERGY TERMS FOR DIPOLE COUPLING, SHAPE ANISOTROPY AND STRESS ANISOTROPY

Let the magnetizations of the two nanomagnets have polar and azimuthal angles of $\theta_1(t)$, $\phi_1(t)$ and $\theta_2(t)$, $\phi_2(t)$, respectively. In that case, the different contributions to the potential energy of the second nanomagnet can be expressed as [19],

$$E_{dipole}(t) = \frac{\mu_0 M_s^2 \Omega^2}{4\pi R^3}[-2(\sin\theta_1(t)\cos\phi_1(t))(\sin\theta_2(t)\cos\phi_2(t)) \quad \text{(A1)}$$
$$+(\sin\theta_1(t)\sin\phi_1(t))(\sin\theta_2(t)\sin\phi_2(t))+\cos\theta_1(t)\cos\theta_2(t)]$$

$$E_{shape\,anisotropy}(t) = \frac{\mu_0 M_s^2 \Omega}{2}[N_{d-xx}(\sin\theta_2(t)\cos\phi_2(t))^2$$
$$+ N_{d-yy}(\sin\theta_2(t)\sin\phi_2(t))^2 + N_{d-zz}(\cos\theta_2(t))^2] \quad \text{(A2)}$$

where $N_{d-xx}$, $N_{d-yy}$ and $N_{d-zz}$ are the demagnetization factors along the x, y and z directions and satisfy the condition

$$N_{d-xx} + N_{d-yy} + N_{d-zz} = 1. \quad \text{(A3)}$$

These equations are valid for both elliptical and cylindrical nanomagnets.

The demagnetization factors for elliptical nanomagnets with major and minor axes diameters of *a* and *b*, and thickness *t* can be approximately expressed [15]. While this approximation is less valid at large (a/b) aspect ratios, Fig. B3 (a) shows that for the dimensions we consider, the approximate macrospin and more accurate micromagnetic simulations show a close match. The demagnetization factors for cylindrical nanomagnets with diameter of $2a$ and length of $2l$ are obtained from [17].

Switching is achieved by applying a uniaxial compressive stress which is along the y axis for the elliptical and along the z axis for the cylindrical nanomagnet. Stress anisotropy energy for the elliptical nanomagnet can be expressed as

$$E_{stress\,anisotropy}(t) = -(3/2)\lambda_s\sigma\Omega(\sin\theta_2(t)\sin\phi_2(t))^2. \quad \text{(A4)}$$

For the cylindrical nanomagnet, the stress is applied along the z axis and the stress anisotropy energy can be expressed as

$$E_{stress\,anisotropy}(t) = -(3/2)\lambda_s\sigma\Omega(\cos\theta_2(t))^2, \quad \text{(A5)}$$

where $(3/2)\lambda_s$ is the saturation magnetostriction and $\sigma$ is the applied uniaxial stress. We can calculate the energies for the second nanomagnet in a similar manner. Since no stress is applied to the first nanomagnet, the stress anisotropy energy is zero.

### II. Analytical expression for the effective field ($H_{eff}$) for elliptical and cylindrical nanomagnets

The quantities $H_{eff-x}$, $H_{eff-y}$ and $H_{eff-z}$ are the x-, y- and z-components of the effective magnetic field $\vec{H}_{eff}$. For the elliptical nanomagnet, they can be expressed as [19]:

$$H^i_{eff-x}(t) = \left(\frac{M_s\Omega}{4\pi R^3}\right)(2\sin\theta_j(t)\cos\phi_j(t))$$
$$- M_s N_{d-xx}\sin\theta_i(t)\cos\phi_i(t) + \sqrt{\frac{2kT\alpha}{\mu_0 M_s \gamma \Omega \Delta t}}(G_x(t))$$

$$H^i_{eff-y}(t) = -\left(\frac{M_s\Omega}{4\pi R^3}\right)(\sin\theta_j(t)\sin\phi_j(t)) - M_s N_{d-xx}(\sin\theta_i(t)\sin\phi_i(t)) \quad \text{(A6)}$$
$$+ \left(\frac{3\lambda_s}{\mu_0 M_s}\right)\sigma_i \sin\theta_i(t)\sin\phi_i(t) + \sqrt{\frac{2kT\alpha}{\mu_0 M_s \gamma \Omega \Delta t}}(G_y(t))$$

$$H^i_{eff-z}(t) = -\left(\frac{M_s\Omega}{4\pi R^3}\right)(\cos\theta_j(t)) - M_s N_{d-xx}\cos\theta_i(t) + \sqrt{\frac{2kT\alpha}{\mu_0 M_s \gamma \Omega \Delta t}}(G_z(t))$$

For the cylindrical nanomagnet, they can be expressed as

$$H_{eff-x}^{i}(t) = \left(\frac{M_s\Omega}{4\pi R^3}\right)\left(2\sin\theta_j(t)\cos\phi_j(t)\right) -$$
$$M_s N_{d-xx}\sin\theta_i(t)\cos\phi_i(t) + \sqrt{\frac{2kT\alpha}{\mu_0 M_S \gamma \Omega \Delta t}}(G_x(t))$$

$$H_{eff-y}^{i}(t) = -\left(\frac{M_s\Omega}{4\pi R^3}\right)\left(\sin\theta_j(t)\sin\phi_j(t)\right) -$$
$$M_s N_{d-xx}\sin\theta_i(t)\sin\phi_i(t) + \sqrt{\frac{2kT\alpha}{\mu_0 M_S \gamma \Omega \Delta t}}(G_y(t))$$

(A7)

$$H_{eff-z}^{i}(t) = -\left(\frac{M_s\Omega}{4\pi R^3}\right)\left(\cos\theta_j(t)\right) - M_s N_{d-xx}\cos\theta_i(t)$$
$$+ \left(\frac{3\lambda_s}{\mu_0 M_s}\right)\sigma_i\cos\theta_i(t) + \sqrt{\frac{2kT\alpha}{\mu_0 M_S \gamma \Omega \Delta t}}(G_z(t))$$

APPENDIX B

Here we show micromagnetic simulations of magnetization dynamics carried out with OOMMF [25] where the effect of applying stress was incorporated by varying the uniaxial anisotropy constant ($K_u$) with a script file. Fig. B1 and Fig. B2 respectively show that the switching behavior of the ellipse and cylinder described in this paper are nearly coherent. Figures Fig. B3 and Fig. B4 respectively show a comparison between the magnetization dynamics computed using macrospin LLG and micromagnetic simulation for the ellipse and cylinder. The close agreement between the two results validates the macrospin approximation used in this paper. Ref [26] also supports our findings qualitatively.

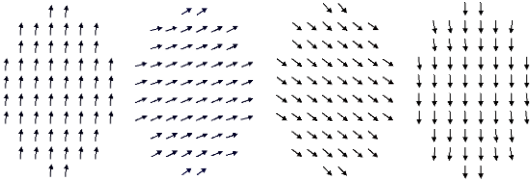

Fig. B1. Results of OOOMF simulation of the switching behavior of an elliptical nanomagnet: different snapshots of spin orientations in time are shown.

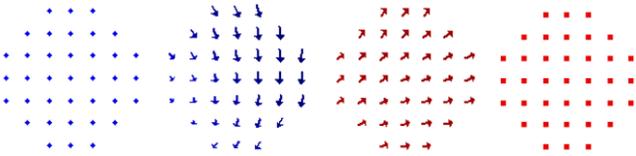

Fig. B2. Results of OOOMF simulation of cylindrical nanomagnet switching (cross section view at center of cylinder). In the far left figure, the blue dots indicate that the magnetization is pointing up along the cylinder axis and in the far right figure the red dots indicate that the magnetization is pointing down. Different snapshots in time are shown.

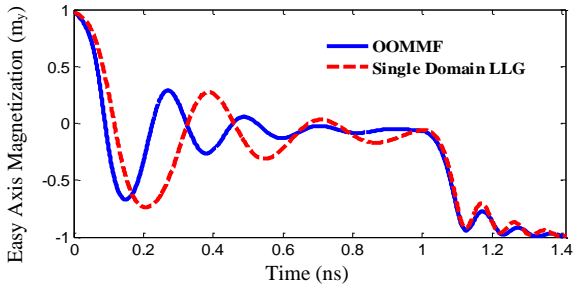

Fig. B3. Comparison between magnetization dynamics predicted by OOMMF and single domain LLG for elliptical nanomagnets. The component of magnetization along the easy axis is plotted versus time.

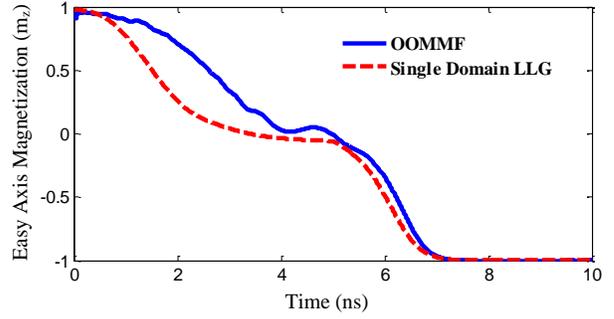

Fig. B4. Comparison between magnetization dynamics predicted by OOMMF and single domain LLG for cylindrical nanomagnets. The component of magnetization along the easy axis is plotted versus time.

ACKNOWLEDGMENT

We acknowledge discussions with Dr. Michael Donahue at NIST, Gaithersburg on the micromagnetic simulations.

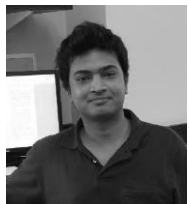
**Md Mamun Al-Rashid** received his B.Sc. degree in Electrical and Electronic Engineering from Bangladesh University of Engineering and Technology, Dhaka, Bangladesh, in 2012. He is currently pursuing his Ph.D. degree in Electrical and Computer Engineering and is also affiliated with the Department of Mechanical and Nuclear Engineering at Virginia Commonwealth University, Richmond, VA, USA.

His current research interests include design, simulation, and fabrication of energy-efficient straintronic multiferroic nanomagnetic devices and modeling of nanoscale stochastic magnetization dynamics.

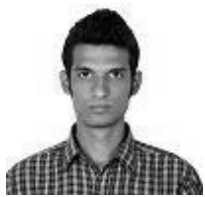
**Dhritiman Bhattacharya** received his B.Sc. degree in Electrical and Electronic Engineering from Bangladesh University of Engineering and Technology, Dhaka, Bangladesh, in 2013. He is currently pursuing his Ph.D. degree in Mechanical and Nuclear Engineering at Virginia Commonwealth University, Richmond, VA, USA.

His current research interests include modeling of nanoscale stochastic magnetization dynamics and multiferroic nanomagnetic devices.

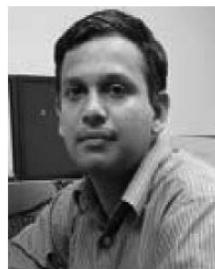
**Jayasimha Atulasimha** (SM'11) received the M.S. and Ph.D. degrees in Aerospace Engineering from the University of Maryland, College Park, MD, USA, in 2003 and 2006, respectively. He is an Associate Professor of Mechanical and Nuclear Engineering and also has a courtesy appointment as Associate Professor of Electrical and Computer Engineering at the Virginia Commonwealth University, Richmond, VA, USA. He has authored or coauthored more than 40 scientific articles including 35 journal publications on magnetostrictive materials, magnetization dynamics, and nanomagnetic computing. His research interests include magnetostrictive materials, nanoscale magnetization dynamics, and multiferroic nanomagnet-based computing architectures.

Prof. Atulasimha received the NSF CAREER Award for 2013–2018 and is a senior member of the IEEE.

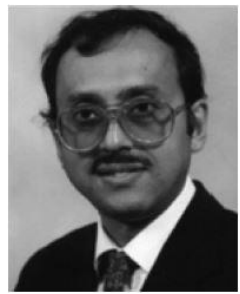
**Supriyo Bandyopadhyay** (SM'80–M'86–SM'88–F'05) is a Professor of Electrical and Computer Engineering at Virginia Commonwealth University, Richmond, VA, USA, where he directs the Quantum Device Laboratory. He has authored and coauthored more than 300 scientific publications. His current research interests include nanoelectronics, nanomagnetism and spintronics.

Prof. Bandyopadhyay chairs the Technical Committee on Spintronics within the IEEE Nanotechnology Council, and is a Fellow the American Physical Society, the Electrochemical Society, the Institute of Physics, and the American Association for the Advancement of Science.